\documentclass[%
 reprint,
 amsmath,amssymb,
 aps,
]{revtex4-2}

\usepackage{graphicx}% Include figure files
\usepackage{dcolumn}% Align table columns on decimal point
\usepackage{bm}% bold math
\usepackage{float}
\usepackage{subfig}
\usepackage{ragged2e}
\usepackage{xcolor}
\usepackage{lipsum}

\newcommand*{\affaddr}[1]{#1} % No op here. Customize it for different styles.
\newcommand*{\affmark}[1][*]{\textsuperscript{#1}}

\begin{document}

\preprint{APS/123-QED}

\title{A Heavy Ion Monitor on a Chip Based on a Non-Volatile Memory Architecture -- Part II: Device Characterization \& Modeling} % Force line breaks with \\

\author{
Dale Julson\affmark[1], Mike Youngs\affmark[2], Hannah Lowrey\affmark[2,3], David Keltner\affmark[2,4], Tim Hossain\affmark[1], Clayton Fullwood\affmark[1]\\
\affaddr{\affmark[1] Cerium Laboratories, Austin, TX, 78741, US}\\
\affaddr{\affmark[2] Texas A$\&$M University Cyclotron Institute, College Station, TX 77843,  USA}\\
\affaddr{\affmark[3] Department of Physics, University of Dallas, Irving, TX 75062,  USA}\\
\affaddr{\affmark[4] Department of Physics and Astronomy, University of Missouri-Kansas City, Kansas City, MO 64110,  USA}\\
}

\date{\today}

\begin{abstract}
Building on the demonstrated sensitivity of the Heavy Ion Monitor on a Chip (HIMoC) presented in Part I of this work, we performed additional irradiation exposures using 24.8 MeV/u beams of $^{14}$N, $^{22}$Ne, and $^{40}$Ar at the Texas A\&M University Cyclotron Institute. A novel simulation workflow was developed that couples the particle-transport toolkit \textit{Geant4} with the open-source TCAD simulator \textit{DEVSIM} to model the heavy-ion-induced signal in HIMoC devices. The model represents energy deposition by primary heavy ions and secondary electrons as Gaussian charge-loss profiles that produce measurable threshold-voltage shifts in the device. Good agreement between simulated and experimental $\Delta V_{\mathrm{th}}$ distributions was obtained. HIMoC was also shown to generate a signal that scales approximately linearly with a dose-like quantity proportional to ion fluence, LET, and active detector area. These results support HIMoC as a passive heavy-ion dosimeter and provide a framework for modeling the effects of radiation-induced charge loss in charge-trapping non-volatile memory devices.
\end{abstract}

\pacs{Valid PACS appear here}% PACS, the Physics and Astronomy
\maketitle

\section{Introduction}

A novel heavy-ion detector technology based on nitride read-only memory (NROM) transistors, referred to as the Heavy Ion Monitor on a Chip (HIMoC), was previously proposed and its sensitivity studied using a commercial off-the-shelf (COTS) memory chip~\cite{HIMoC_Paper}. In this work, we build on that prior study by evaluating the HIMoC response to additional heavy-ion species across varying beam energies and exposure fluences. We also develop an open-source simulation framework for modeling the post-irradiation device response, providing a pathway to better understand the relationship between heavy-ion interactions and the measured electrical signal. A key feature of HIMoC is its ability to record heavy-ion interactions without requiring active power during irradiation. This passive detection capability makes the technology well suited for environments where power availability, system complexity, or long-duration deployment are important design constraints, for instance in space-based heavy-ion monitoring and dosimetry.

The paper is organized as follows. Section~II provides background on existing radiation monitoring technologies and nonvolatile-memory-based radiation detectors. Section~III describes the experimental setup used in this study. Section~IV presents the simulation methodology used to model the induced HIMoC signal following heavy-ion exposure. Section~V presents the experimental measurements, compares them with the simulation results, and provides additional discussion. Section~VI summarizes the conclusions and next steps, and Section~VII provides acknowledgments.

\section{Background}

\begin{figure*}[t]
  \centering
  \raisebox{7mm}{\includegraphics[width=0.43\textwidth]{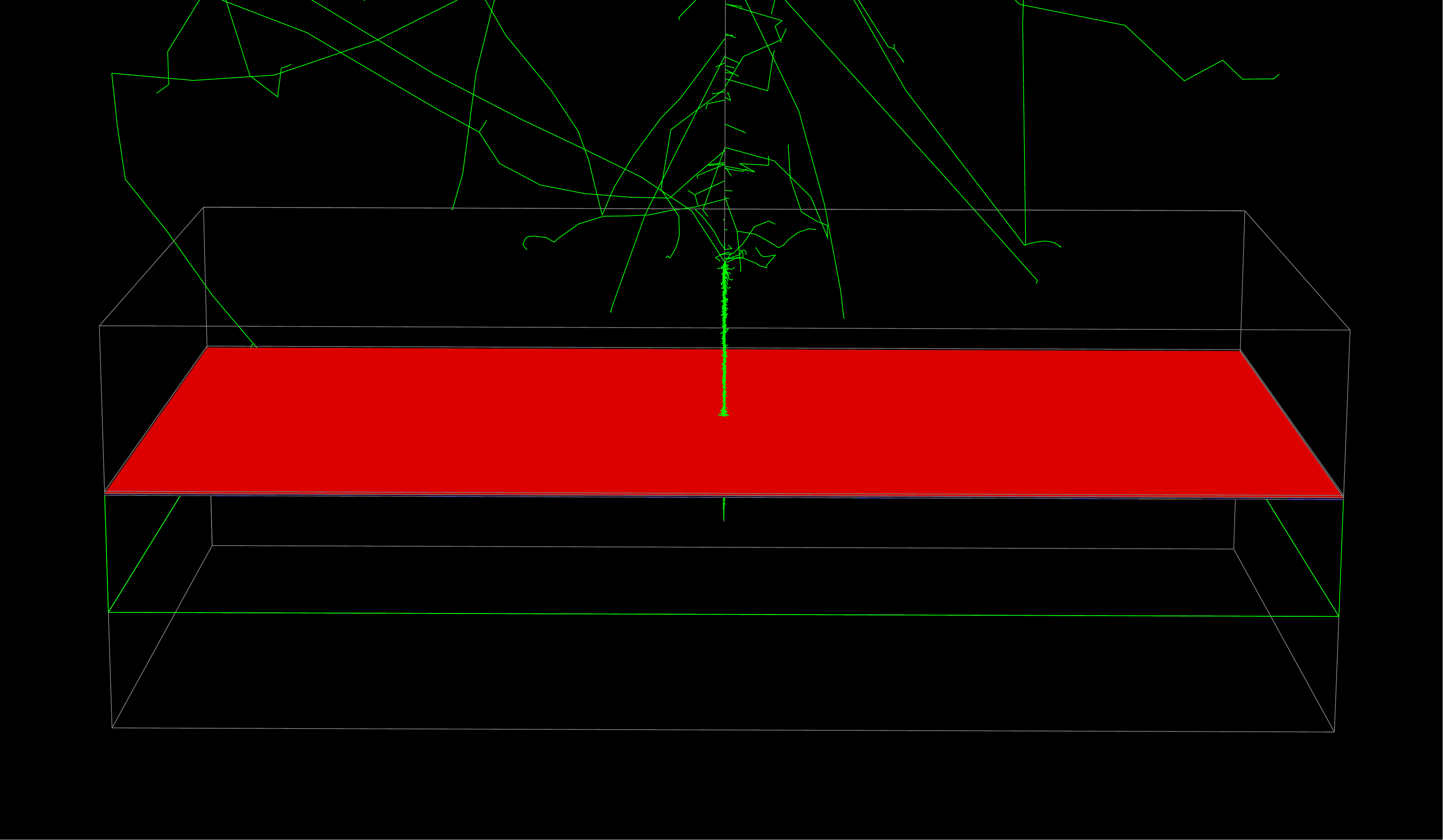}}
  \hspace{1cm}
  \includegraphics[width=0.43\textwidth]{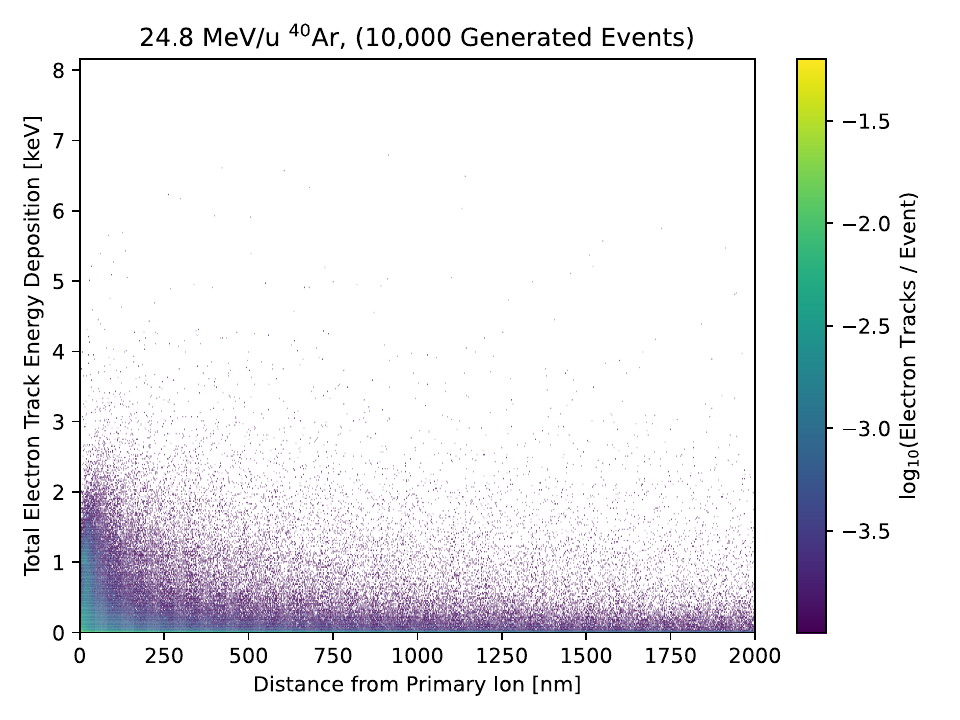}
  \caption{\parbox{0.95\textwidth}{\justifying
  (Left) Geant4 visualization of a 24.8 MeV/u $^{40}$Ar ion impinging on HIMoC. The modeled device geometry is approximately 3 mm in length and width, and 1 mm in depth. The outlined gray boxes represent the plastic packaging, the outlined green box represents the Si substrate, and the red regions represent metal interconnect and dielectric layers. Secondary electrons are shown in green. (Right) Logarithmic map of the energy deposited by secondary electrons in the detector sensitive region as a function of distance from the primary heavy ion, also for 24.8 MeV/u $^{40}$Ar.}}
  \label{fig:Ar40}
\end{figure*}

The ability to accurately measure and record total received radiation dose is critical for human safety in environments such as nuclear power generation, radiation-based cancer therapy, and human space exploration. Among these environments, space is particularly challenging because of the diversity of radiation species, the large variation in possible dose rates, and the potential for abrupt increases in radiation exposure driven by solar energetic particle events and other space-weather phenomena~\cite{radiation}. These challenges are compounded by strict constraints on detector power consumption and mass, as well as by the need for timely dose information without requiring post-exposure processing on Earth that could delay treatment for radiation-sickened individuals. These considerations have motivated recent interest in low-power, low-mass radiation monitoring technologies capable of real-time readout. Examples include the International Space Station Radiation Assessment Detector (ISS-RAD), the Hybrid Electronic Radiation Assessor (HERA), and NASA's Crew Active Dosimeter~\cite{ISS_RAD, time_pix, crew_active_dosimeter}. 

In parallel, particle detector technologies based on transistor structures commonly used in non-volatile memory (NVM) devices have recently been proposed and evaluated for applications including heavy-ion and neutron detection~\cite{HIMoC_Paper, NAND_Heavy_Ion_Detector, flash_detector, NISoC, Tower_detector}. Of particular interest here is HIMoC, a flash-memory-based heavy-ion detector implemented in a radiation-hardened technology node, making it well suited for space radiation monitoring. The sensing elements in flash-based detectors resemble conventional MOSFETs, but include an additional charge-trapping layer capable of storing electric charge located inbetween a \textit{gate oxide} layer and \textit{tunnel oxide} layer. During a write process, charge is injected from the substrate into this layer through hot-carrier injection or Fowler-Nordheim tunneling. For an NMOS-style device, electron injection increases the transistor threshold voltage, $V_{\mathrm{th}}$.

The sensing procedure begins by writing charge into the detector array, which can contain more than $2^{30}$ sensing elements. This write operation increases the $V_{\mathrm{th}}$ of each programmed element. When the device is subsequently exposed to heavy-ion radiation, trapped electrons can be discharged from the charge-trapping layer~\cite{heavy_ion_FG_effects}, producing a negative shift in $V_{\mathrm{th}}$ back toward its pre-programmed value. The magnitude and spatial distribution of these threshold-voltage shifts provide the basis for particle detection and dosimetry.

For HIMoC, the charge-trapping layer is composed of the dielectric $\mathrm{Si}_{3}\mathrm{N}_{4}$. This differs from many other flash-based detectors, which use a conductive polysilicon floating gate (FG). The distinction is important because $\mathrm{Si}_{3}\mathrm{N}_{4}$ has been shown to avoid some of the deleterious long-term effects observed in FG devices after heavy-ion exposure~\cite{Rad_effects_NROM}. In FG devices, charge can freely redistribute across the equipotential floating-gate layer. As a result, a defect induced by a heavy ion in the tunnel oxide between the FG and substrate can lead to complete discharge of a sensing element, including continued discharge long after the radiation exposure has ended~\cite{data_retention}. In contrast, the localized charge storage in the nitride layer of NROM devices can mitigate this failure mode. HIMoC also benefits from the ability of a single transistor to store two independently readable charge packets, referred to here as the \textit{two-bit effect}~\cite{NROM}. This increases the effective density of sensing elements and can improve sensitivity, as discussed in Section~\ref{sec:Modeling}.

When energetic heavy ions pass through dielectric layers in modern semiconductor devices, they generate dense, approximately columnar distributions of electron-hole pairs~\cite{recombination}. Most of these carriers quickly recombine, leaving only a small fraction of mobile holes and electrons available to contribute to trapped-charge loss~\cite{charge_yield}. These charge distributions are often modeled as approximately Gaussian in the transverse direction~\cite{gaussian_shape}. However, the microscopic mechanism responsible for discharging trapped charge remains an active topic of discussion. Proposed mechanisms include \textit{trap-assisted tunneling}, in which radiation-induced oxide defects allow trapped electrons to tunnel out; the \textit{transient conductive path} model, in which the dense electron-hole plasma forms a temporary conductive path to ground; \textit{transport in oxide} models, in which radiation-generated holes recombine with trapped electrons; and \textit{electron emission}, in which trapped electrons gain sufficient energy to escape from localized trap states~\cite{discharge_models}. In this work, we do not explicitly model the microscopic discharge mechanism. Instead, the simulations are based on the final spatial profile of lost trapped charge. Nevertheless, the observed behavior provides useful qualitative guidance. Trap-assisted tunneling and transient conductive path models appear less likely to dominate the HIMoC response because long-term percolative discharge has not been observed in NROM devices, and because the $\mathrm{Si}_{3}\mathrm{N}_{4}$ trapping layer is not conductive in the same manner as a floating gate. In contrast, accurate modeling of secondary-electron effects was found to be necessary for agreement between experimental data and simulation, suggesting that electron-mediated discharge processes may play an important role in the HIMoC response.
 
\section{Experimental Setup}

Heavy-ion exposures of HIMoC devices were performed at the Texas A\&M University Cyclotron Institute using three ion species: 24.8 MeV/u $^{14}$N, 24.8 MeV/u $^{22}$Ne, and 24.8 MeV/u $^{40}$Ar. A new HIMoC device was used for each irradiation, and no device was reused after exposure. Before irradiation, each device was programmed by writing all bits to logical ``0'', corresponding to charge stored in the charge-trapping region. The device was then read out by sweeping the gate voltage, $V_{\mathrm{g}}$, of each individual bit in 100 mV increments and recording the value at which the bit transitioned from logical ``0'' to logical ``1''. The logical state of each bit is determined internally by sense-amplifier comparators, which compare the drain current, $I_{\mathrm{d}}$, to a predetermined threshold. Because $I_{\mathrm{d}}$ depends directly on the difference between $V_{\mathrm{g}}$ and the transistor threshold voltage, $V_{\mathrm{th}}$, the gate voltage at which the logical transition occurs provides a measure of $V_{\mathrm{th}}$ for that bit. The same readout procedure was repeated after irradiation. The threshold-voltage shift, $\Delta V_{\mathrm{th}}$, was then calculated on a bit-by-bit basis as the difference between the pre- and post-irradiation readout values.

The $^{14}$N exposure was performed at a fluence of $2\times10^{8}~\mathrm{cm}^{-2}$ over an exposure time of $4.46\times10^{3}$ s. Two $^{22}$Ne exposures were performed, each at a fluence of $2\times10^{8}~\mathrm{cm}^{-2}$. One device was exposed for $7.49\times10^{3}$ s, referred to as the \textit{low-flux} exposure, while the other was exposed for $2.23\times10^{2}$ s, referred to as the \textit{high-flux} exposure. The $^{40}$Ar exposure was performed at a fluence of $1.7\times10^{8}~\mathrm{cm}^{-2}$ over an exposure time of $2.99\times10^{3}$ s. The data recorded from these exposures are presented in Section~\ref{sec:Results}.

A readout-induced noise contribution was also observed and attributed to the sense amplifiers internal to the device. As a result, even an unirradiated device exhibits a small, approximately Gaussian distribution of apparent $\Delta V_{\mathrm{th}}$ values when read out repeatedly. The mean, $\mu$, and standard deviation, $\sigma$, of this read-noise distribution were measured by reading out the same device five times and calculating $\Delta V_{\mathrm{th}}$ between consecutive readouts. These results are also presented in Section~\ref{sec:Results}.

\begin{figure*}[t]
  \centering
  \includegraphics[width=0.30\textwidth]{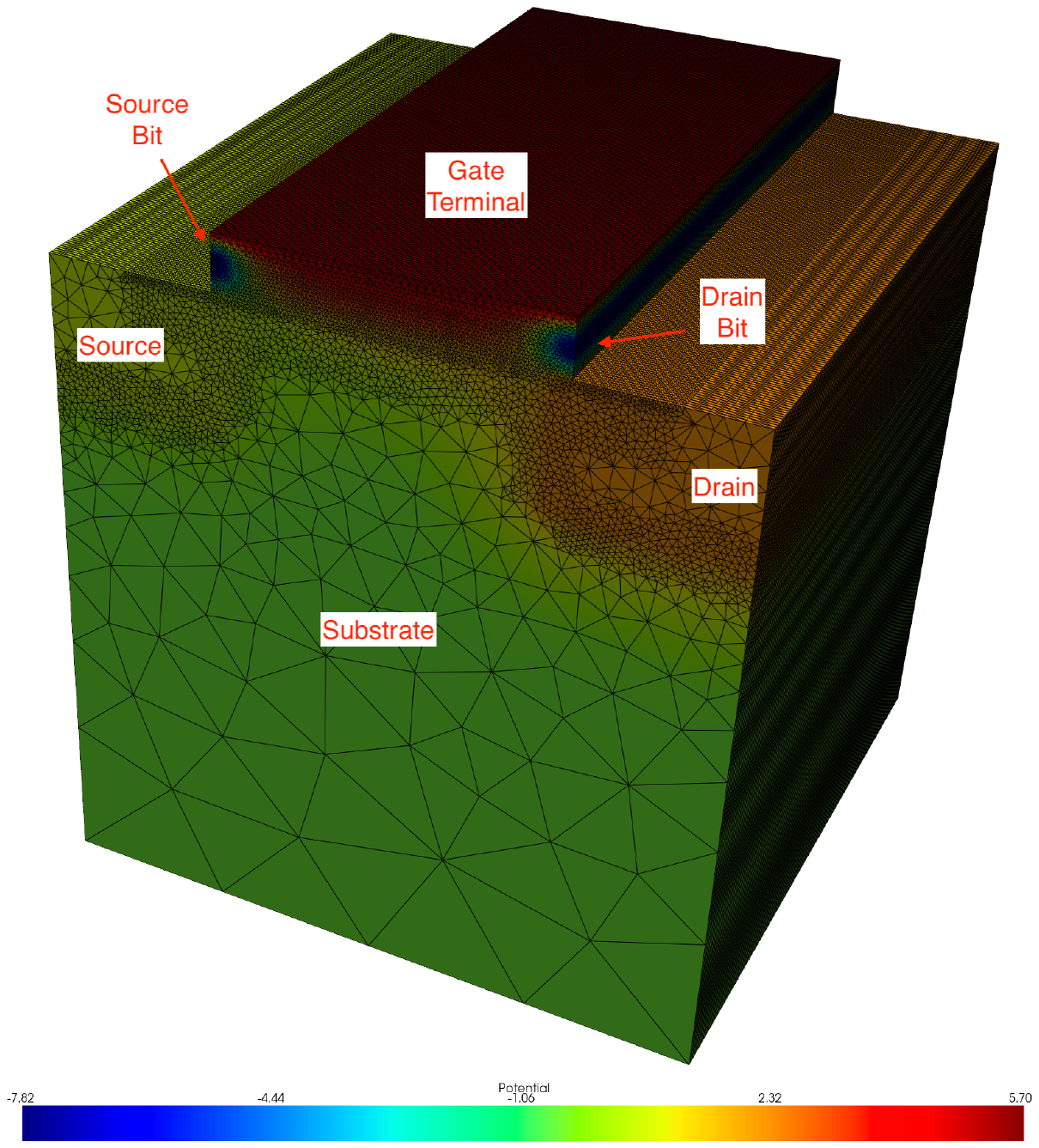}
  % \hfill
  \hspace{1cm}
  \includegraphics[width=0.43\textwidth]{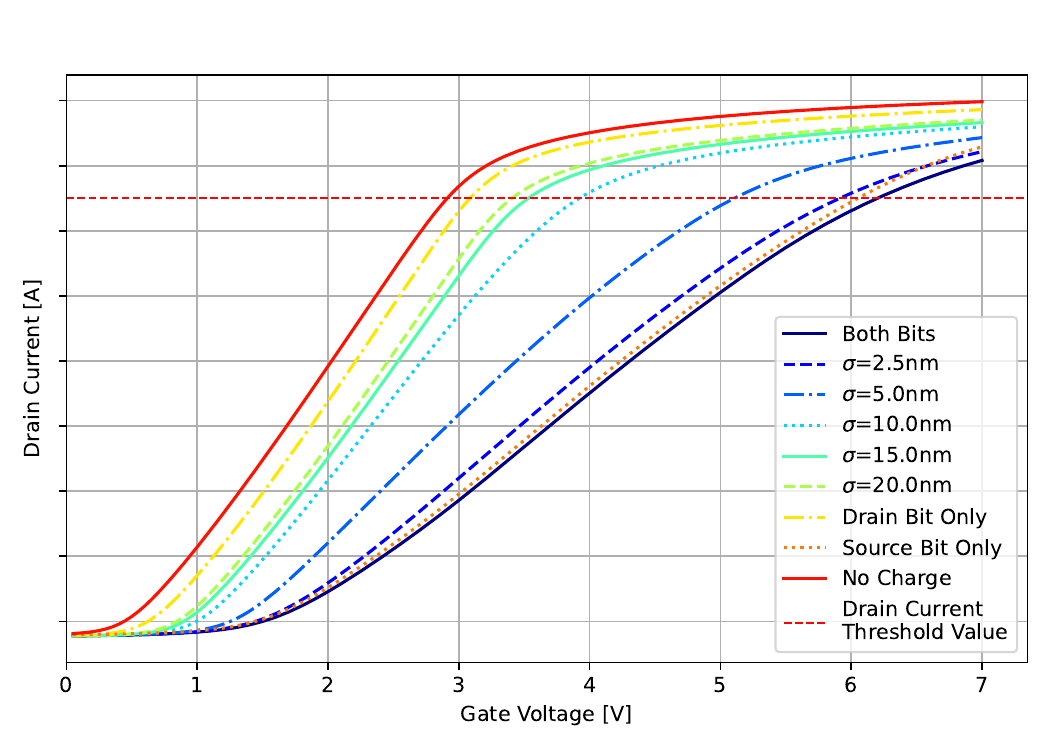}
  \caption{\parbox{0.95\textwidth}{\justifying
  (Left) Three-dimensional NROM transistor mesh with the simulated electric potential overlaid after a full $V_{\mathrm{d}}$ and $V_{\mathrm{g}}$ ramp process in DEVSIM. (Right) The DEVSIM-derived $V_{\mathrm{g}}$--$I_{\mathrm{d}}$ relationship at constant $V_{\mathrm{d}}$ for varying charge-loss scenarios parameterized by $\sigma$ (see Eq.~\ref{eq:gaussian}). This plot demonstrates the two-bit effect: charge present in both bits and charge present only in the source bit produce the same $V_{\mathrm{g}}$--$I_{\mathrm{d}}$ relationship, while no charge present in either bit and charge present only in the drain bit also produce the same $V_{\mathrm{g}}$--$I_{\mathrm{d}}$ relationship. Y-axis values have been intentionally omitted.}}
  \label{fig:DEVSIM}
\end{figure*}

\section{Modeling \& Simulations}\label{sec:Modeling}

To better understand the response of HIMoC devices to heavy-ion radiation, we developed a simulation workflow using open-source software tools. The workflow models both the radiation transport through the device under test, including primary-ion interactions, secondary-particle generation, and energy deposition, as well as the resulting response of each individual sensing element after irradiation.

The simulation workflow consists of two main steps. First, primary-ion and secondary-particle energy-deposition profiles are simulated using the particle-tracking toolkit \textit{Geant4}~\cite{GEANT1, GEANT2, GEANT3}. Second, the particle-induced energy deposition is mapped onto a device-response model derived from TCAD simulations. This response model is applied on a per-sensing-element basis to estimate the resulting threshold-voltage shift, $\Delta V_{\mathrm{th}}$. The simulated $\Delta V_{\mathrm{th}}$ distributions are then compared with the experimental data, and free parameters of the model are iteratively adjusted to improve agreement between simulation and measurement.

\subsection{Geant4 Modeling}

Two effects must be accurately represented to model the HIMoC response to heavy-ion irradiation. First, as discussed above, heavy ions generate large numbers of secondary electrons in semiconductor devices. The secondary-electron multiplicity must therefore be modeled accurately. Second, the sensitive regions in HIMoC are extremely thin, with characteristic thicknesses of approximately 5--15 nm. Accurate modeling of energy deposition within these thin regions is therefore also required.

To address these requirements, the Geant4 simulations were performed in two stages. The first stage was used to estimate the multiplicity of secondary electrons produced per incident heavy ion. For physics lists such as the \textit{Geant4 Low Energy Electromagnetic} package, secondary-electron production thresholds are generally not recommended below 250 eV~\cite{Geant4_Low_Energy_EM}. This threshold is much larger than the reported trap depth in $\mathrm{Si}_{3}\mathrm{N}_{4}$ charge trapping layers which is approximately 1.4 eV below the conduction band~\cite{SiN_trap_depth}. We therefore used the \textit{Geant4-MicroElec} physics package, which can model secondary-electron production to significantly lower energies~\cite{Geant4_MuElec}. The main limitation of the \textit{Geant4-MicroElec} package is that it supports only a limited subset of the materials present in HIMoC. Therefore, for the secondary-electron multiplicity calculation, the sensitive region was approximated as a 15 nm thick Si target located beneath 4 $\mu$m of Si. This simplified geometry approximately matches the relevant depth scale of the HIMoC sensitive region while remaining compatible with the available material models. Primary ions were propagated through the Si target, and the number of secondary electrons entering the target region was recorded on a per-event basis. This produced a distribution of secondary electrons per incident ion. The secondary-electron production threshold was treated as an effective model parameter, and a value of 8.095 eV was used in the final simulation set as it was found to provide the best agreement between simulation and experiment.

Although this first stage provides an estimate of secondary-electron multiplicity, HIMoC is not composed only of Si but rather the full device includes multiple dielectric layers, metal interconnects, and exterior plastic packaging. A second Geant4 stage was therefore used to model energy deposition in the complete device geometry. For this stage, the \textit{Geant4 EM Opt4} physics package was used. Figure~\ref{fig:Ar40} (Left) shows a visualization of a 24.8 MeV/u $^{40}$Ar ion impinging on the device, while Figure~\ref{fig:Ar40} (Right) shows the corresponding two-dimensional map of secondary-electron energy deposition as a function of distance from the primary ion. Similar two-dimensional energy-deposition maps were generated for the 24.8 MeV/u $^{14}$N and 24.8 MeV/u $^{22}$Ne beams. One-dimensional distributions of primary-ion energy deposition were also produced for each ion species.

The two Geant4 stages are combined in the full simulation workflow as follows. The secondary-electron multiplicity distribution is sampled to determine the number of secondary electrons associated with each incident heavy ion to be generated. The energy-versus-distance maps are then sampled to assign each secondary electron a radial position relative to the primary-ion track and an energy deposition. These sampled electron contributions are then propagated into the device-response model, as described below. Finally, separate Geant4 simulations were performed to estimate the linear energy transfer (LET) of each primary ion as it traversed the sensitive volume. In order to ensure that the extracted LET values were not dominated by artifacts associated with the small sensitive-volume thickness, simulations were repeated using sensitive-volume thicknesses of 1 $\mu$m, 0.5 $\mu$m, and 0.1 $\mu$m. The resulting LET values were verified to converge to a consistent value, and results are presented in table~\ref{tab:parameters}.

\begin{figure*}[t]
  \centering
  \includegraphics[width=0.23\textwidth]{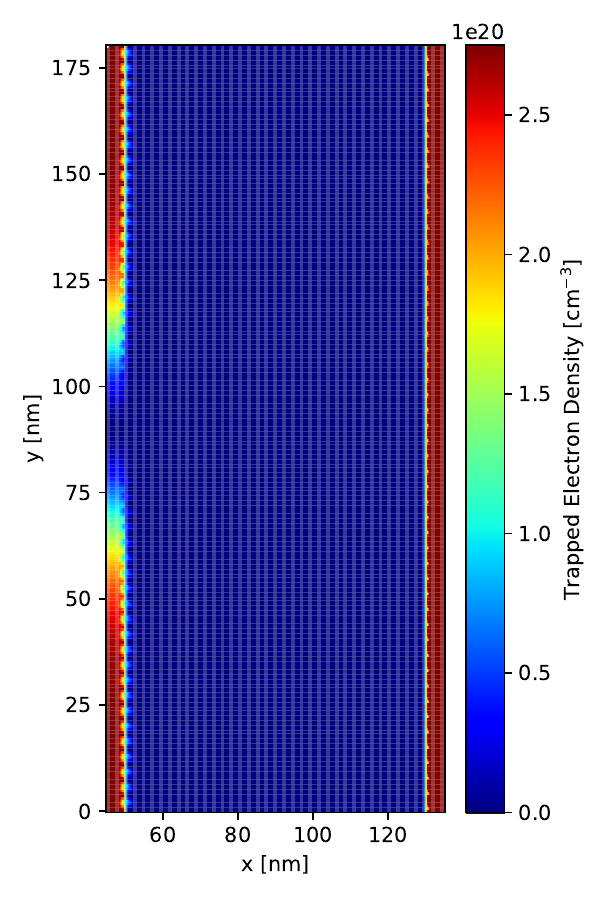}
  % \hfill
  \hspace{1cm}
  \includegraphics[width=0.43\textwidth]{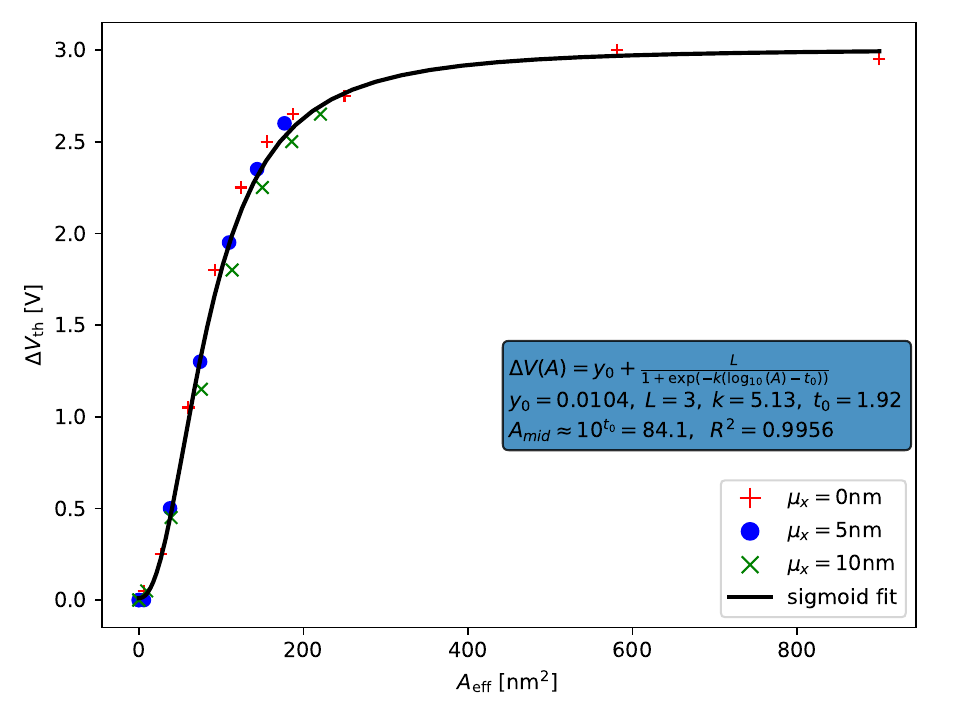}
  \caption{\parbox{0.95\textwidth}{\justifying
  (Left) Two-dimensional $xy$ profile of the Si$_{3}$N$_{4}$ charge-trapping region showing the trapped electron density after Eq.~\ref{eq:rho} has been applied to the source bit for a charge-loss profile with $\sigma=20~\mathrm{nm}$ and $\mu_x=\mu_y=0~\mathrm{nm}$. (Right) Induced $\Delta V_{\mathrm{th}}$ as a function of $A_{\mathrm{eff}}$ (see Eq.~\ref{eq:effective_area}). The data points correspond to Gaussian charge-loss profiles with varying $\sigma$ values and different center positions along the $x$ direction, $\mu_x=0~\mathrm{nm}$, $5~\mathrm{nm}$, and $10~\mathrm{nm}$. The black curve is a sigmoid fit to the simulated data, with the fit parameters shown in the blue text box.}}
  \label{fig:Effects}
\end{figure*}

\subsection{TCAD Modeling}

The second component of the simulation workflow involved deriving a device-response relationship between charge loss in the charge-trapping region and the resulting threshold-voltage shift, $\Delta V_{\mathrm{th}}$. Although many commercial software packages can simulate semiconductor and MOSFET devices, such simulations are often restricted to two-dimensional cross sections along the channel length and depth, or to ``2.5-dimensional'' simulations in which symmetry is assumed along the device width. For HIMoC devices exposed to heavy-ion irradiation, there is no reason to expect the trapped-charge distribution in the nitride layer to remain symmetric along the device width. Instead, highly localized and irregular charge-loss configurations are expected. The open-source TCAD semiconductor device simulator \textit{DEVSIM} was therefore used because of its demonstrated capability for full three-dimensional semiconductor device simulation~\cite{DEVSIM1, DEVSIM2}.

Figure~\ref{fig:DEVSIM} (Left) shows the three-dimensional mesh used to simulate the relationship between applied gate voltage and drain current, $V_{\mathrm{g}}$--$I_{\mathrm{d}}$, which underlies the HIMoC sensing mechanism. Simulated trapped-charge configurations were placed within the thin regions labeled ``Source Bit'' and ``Drain Bit'' in the mesh. Importantly, the model was able to reproduce the \textit{two-bit effect}, in which the threshold-voltage shift of each bit can be measured independently by switching the applied source voltage ($V_{\mathrm{s}}$) and drain voltage ($V_{\mathrm{d}}$). After confirming that the simulated unexposed device produced the expected electrical behavior, charge-loss configurations were introduced into the charge-trapping layer to determine the relationship between lost trapped charge and $\Delta V_{\mathrm{th}}$.

In a conventional MOSFET, $V_{\mathrm{th}}$ is largely unaffected by radiation exposure except in cases involving permanent damage mechanisms, such as single-event dielectric rupture or trap generation in the gate oxide after large total ionizing doses~\cite{SEDR, TID_effects}. In FG devices, the relationship between total charge loss in the floating gate, $\Delta Q$, and the resulting threshold-voltage shift is commonly written as $\Delta Q = C_{\mathrm{Tot}}\Delta V_{\mathrm{th}}$, where $C_{\mathrm{Tot}}$ is the total capacitance of the floating gate~\cite{discharge_models}. This relationship follows from the conductive nature of the floating gate, which allows charge to redistribute across an equipotential surface. As a result, the threshold-voltage shift is primarily determined by the total charge lost rather than by the detailed spatial distribution of that loss.

For NROM devices, and dielectric based charge-trapping devices more generally, this simple relationship no longer holds~\cite{Sadd_paper}. Because charge is stored locally in the nitride layer, the spatial distribution of charge loss can strongly influence the measured $\Delta V_{\mathrm{th}}$. In fact, for sufficiently non-homogeneous charge trapping configurations, the application of $V_{\mathrm{g}}$ can cause certain regions of the substrate to enter into strong inversion while other regions remain in weak inversion~\cite{Modeling_Nanocrystal, percolation_path}. The result for NROM based devices is that because the trapped charge is already concentrated in thin regions above the source and drain, localized charge loss can produce large threshold-voltage shifts even when the total lost charge is small. Figure~\ref{fig:DEVSIM} (Right) shows the simulated $V_{\mathrm{g}}$--$I_{\mathrm{D}}$ response at fixed drain voltage for several trapped-charge configurations, including full charge stored in both bits, charge stored only in the source-side bit, charge stored only in the drain-side bit, and several localized charge-loss configurations within the source bit. Because of the two-bit effect, charge loss in the drain-side bit does not affect the source-side readout configuration considered here. The $\sigma$ values in the figure legend describe the effective size of the induced charge loss, as described below.

The relationship between energy deposition in the charge-trapping layer and the resulting charge loss was modeled using an approach similar to that of Ref.~\cite{Modeling_Nanocrystal}. Specifically, the local fractional charge loss produced by a deposited-energy event was modeled as a Gaussian profile,

\begin{equation}\label{eq:gaussian}
    g(x,y)=\exp\!\left[-\frac{(x-\mu_{x})^2+(y-\mu_{y})^2}{2\sigma^2}\right],
\end{equation}

\noindent where a right-handed coordinate system is used with $x$ along the device channel, $y$ along the device width, and $z$ directed from the substrate toward the gate. The parameters $\mu_x$ and $\mu_y$ define the center of the charge-loss profile, and $\sigma$ controls its radial extent. In this model, $g(x,y)$ represents the local fraction of trapped charge removed by the event, with $0 \leq g(x,y) \leq 1$. After energy deposition, the remaining trapped-charge density is written as

\begin{equation}\label{eq:rho}
    \rho(x,y)=\rho_0\left[1-g(x,y)\right],
\end{equation}

\noindent where $\rho_0$ is the initial trapped-charge density of the programmed bit. For all simulations, the charge loss is assumed to be uniform through the thickness of the charge-trapping layer. The total charge lost from an individual bit is then

\begin{equation}
    Q_{\mathrm{loss}} = \rho_0 A_{\mathrm{eff}} t_{\mathrm{nitride}},
\end{equation}

\noindent where

\begin{equation}\label{eq:effective_area}
    A_{\mathrm{eff}}=\iint_D g(x,y)\,dx\,dy.
\end{equation}

\noindent Here, $D$ is the two-dimensional domain spanned by the bit in the $x$ and $y$ directions, and $t_{\mathrm{nitride}}$ is the thickness of the charge-trapping layer. The quantity $A_{\mathrm{eff}}$ is the spatial integral of the fractional charge-loss profile over the bit area, and therefore provides a single scalar measure of the total charge removed by the Gaussian charge-loss distribution. Figure~\ref{fig:Effects} (Left) shows a two-dimensional $xy$ view of the nitride layer after applying this charge-loss model. The drain-side bit, shown in red on the right, remains unaffected by the deposited energy. The source-side bit, shown on the left, has undergone localized charge loss described by the Gaussian profile. Figure~\ref{fig:Effects} (Right) shows the resulting $\Delta V_{\mathrm{th}}$ shift parameterized by $A_{\mathrm{eff}}$, where larger values of $A_{\mathrm{eff}}$ correspond to greater charge loss in the charge-trapping region. To extract this relationship, simulations were performed for Gaussian charge-loss profiles centered at different positions along the $x$ and $y$ directions. Red plus signs indicate profiles centered within the source-side bit, blue circles indicate profiles shifted by 5 nm in the positive $x$ direction, and green exes indicate profiles shifted by 10 nm in the positive $x$ direction. Varying the Gaussian center along either the $x$ or $y$ direction did not produce additional nonlinear dependence beyond that captured by $A_{\mathrm{eff}}$. A sigmoid function, shown in black, was therefore fit to the simulated data and used as the extracted relationship between $\Delta V_{\mathrm{th}}$ and $A_{\mathrm{eff}}$.

\subsection{Combined Workflow}

The HIMoC devices tested in this study are 1Gb NROM NOR flash devices fabricated in a 90 nm process node technology. Directly simulating every bit would be computationally impractical and therefore the simulation was performed on a representative subset of the chip, and the resulting distributions were scaled to the corresponding experimentally analyzed bit population. This reduction decreases computational cost while preserving the expected response for the majority of the distribution. However, it also reduces the number of simulated rare events, particularly cells struck by unusually large numbers of ions. Because these high-multiplicity events occur only in the far tail of the Poisson-style hit distribution, their impact on the overall simulated response of the entire device is expected to be small. Even after this reduction however, simulating all bits simultaneously still remains impractical. The simulated portion of the chip was therefore divided into \textit{sectors}, each containing a $64\times64$ grid of cells, with two independently readable bits per cell. A total of 4096 sectors were simulated. This sector-based approach also enables straightforward parallelization, substantially reducing the total simulation runtime.

The combined workflow proceeded as follows. First, ion hit locations were sampled randomly within the active area $A_{\mathrm{active}}$, defined as the total active area of all cells in the device being simulated. For a given exposure fluence, $\Phi_{\mathrm{tot}}$, the expected number of incident ions in the simulated active area is given by

\begin{equation}
    N_{\mathrm{ion}} = \Phi_{\mathrm{tot}} A_{\mathrm{active}}.
\end{equation}

\noindent The corresponding number of ion hit locations was then generated within this area. After the ion locations were sampled, each sector was simulated individually by placing the relevant ion hits into the corresponding $64\times64$ cell grid in which they fell.

For each incident ion, the secondary-electron multiplicity distribution derived from the Geant4 simulations was sampled to determine the number of secondary electrons to be generated by that ion. For each secondary electron, the radial distance from the parent ion and deposited energy were sampled from the normalized energy-versus-distance maps described above. The azimuthal angle was sampled uniformly over $[0,2\pi]$. Because secondary electrons can travel up to approximately $2~\mu$m from the primary-ion track (and perhaps even further), ion hits and secondary electrons originating in adjacent sectors were also included when they could contribute energy deposition within the sector of interest. Once the locations and deposited energies of all primary ions and secondary electrons were determined, the charge-loss response was evaluated cell by cell. Each cell was discretized into a fine grid with spacing $0.05~\mathrm{nm}\times0.05~\mathrm{nm}$. For cells affected by multiple particles, the individual Gaussian charge-loss profiles were combined into a single loss profile, $G(x,y)$, using the union-like expression

\begin{equation}\label{gaussian_product}
    G(x,y)=1-\prod_i \bigl(1-g_i(x,y)\bigr),
\end{equation}

\noindent where $i$ indexes the particles contributing to charge loss within the cell, and $g_i(x,y)$ is the Gaussian profile defined in Eq.~\ref{eq:gaussian}. For multiple-particle events, $G(x,y)$ replaces $g(x,y)$ in Eq.~\ref{eq:effective_area}. This construction ensures that $0 \leq G(x,y) \leq 1$ for all $(x,y)\in D$. As a result, separated or weakly overlapping Gaussian profiles contribute approximately additively, while strongly overlapping profiles saturate rather than producing unphysical charge loss greater than the initially stored charge.

A relationship is also required between the deposited energy of each charged particle, $E_{\mathrm{dep}}$, and the radial extent, $\sigma$, of the corresponding Gaussian charge-loss profile. The following empirical relationships were used for electrons and heavy ions:

\begin{equation}
\begin{aligned}
\text{Electrons:}\qquad
\sigma_e &= a_e + b_e \log_{10}\!\left(\frac{E_{\mathrm{dep}} + c_e}{1~\mathrm{keV}}\right), \\
\text{Heavy ions:}\qquad
\sigma_i &= a_i + b_i \log_{10}\!\left(\frac{E_{\mathrm{dep}} + c_i}{1~\mathrm{keV}}\right).
\end{aligned}
\end{equation}

\noindent For all models considered, $a_e$ and $a_i$ were set to 0, while $c_e$ and $c_i$ were set to $1~\mathrm{keV}$. With these choices, $\sigma=0$ when $E_{\mathrm{dep}}=0$, and $\sigma>0$ when $E_{\mathrm{dep}}>0$. The factor of $1~\mathrm{keV}$ in the logarithm makes the argument dimensionless, and $\sigma_e$ and $\sigma_i$ are expressed in units of nm. The remaining free parameters, $b_e$ and $b_i$, were used to independently calibrate the electron and heavy-ion responses for each model. After $G(x,y)$ was determined for a given bit, Eq.~\ref{eq:effective_area} was evaluated numerically by approximating the double integral as a sum over the fine spatial grid. The resulting $A_{\mathrm{eff}}$ value was then passed through the TCAD-derived sigmoid response function shown in Fig.~\ref{fig:Effects} to obtain the simulated per-bit threshold-voltage shift, $\Delta V_{\mathrm{th}}$.

\begin{figure*}[t]
  \centering
  \includegraphics[width=0.43\textwidth]{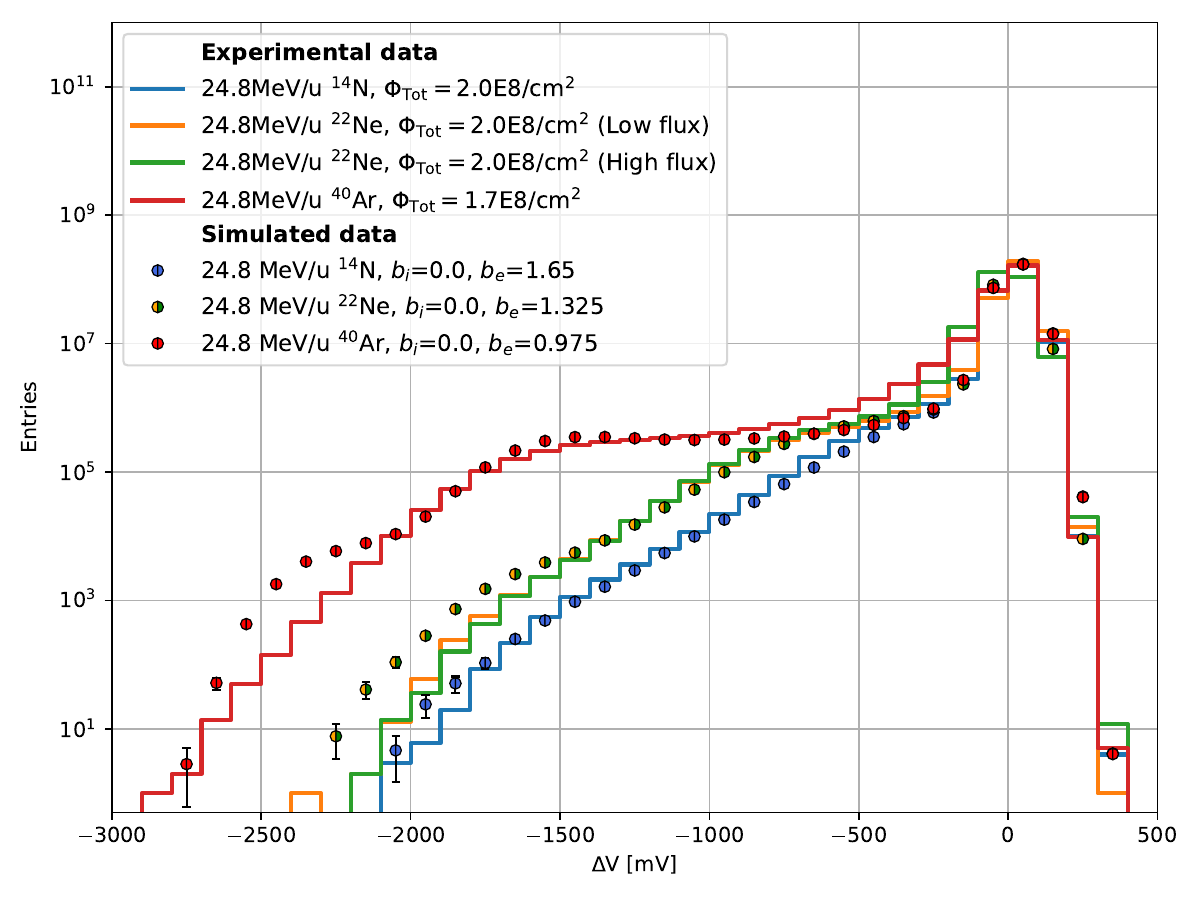}
  % \hfill
  \hspace{1cm}
  \includegraphics[width=0.43\textwidth]{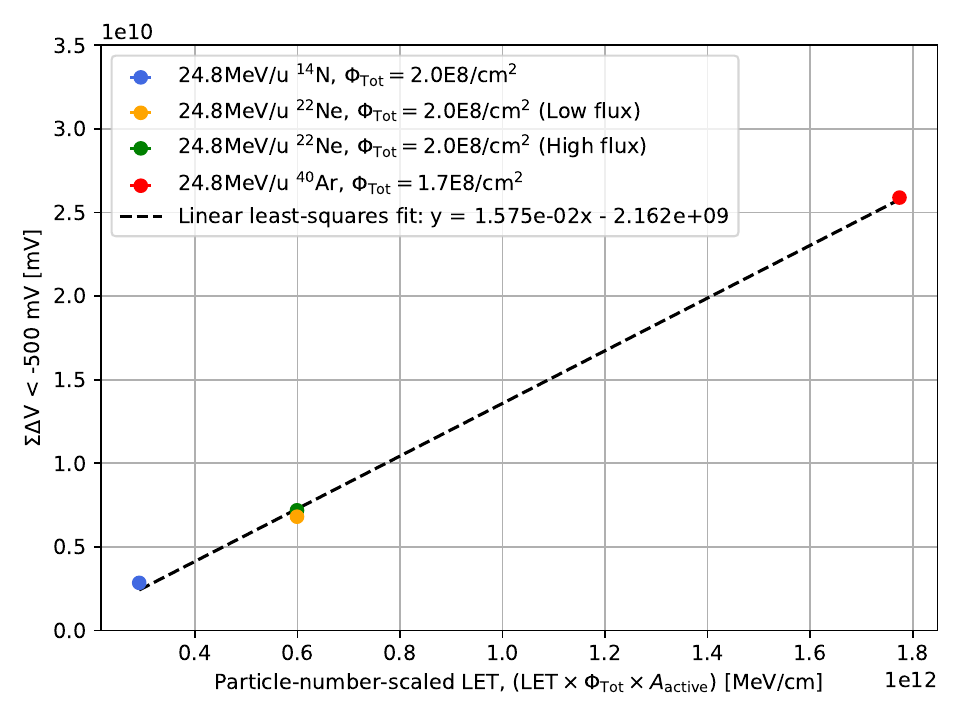}
  \caption{\parbox{0.95\textwidth}{\justifying
  (Left) Measured and simulated $\Delta V_{\mathrm{th}}$ distributions for the HIMoC devices exposed to 24.8 MeV/u $^{14}$N, $^{22}$Ne, and $^{40}$Ar ions. Solid lines indicate experimental data, while markers indicate simulated results. (Right) Summed threshold-voltage shift, $\Sigma\Delta V_{\mathrm{th}}$, for each exposure as a function of the product of ion fluence, mean LET, and active detector area. The cutoff value used to define signal bits was $C_V=-500~\mathrm{mV}$.}}
  \label{fig:Results}
\end{figure*}

\section{Results \& Discussion}\label{sec:Results}

\subsection{Measured and Simulated HIMoC Response}

Figure~\ref{fig:Results} (Left) shows the measured and simulated $\Delta V_{\mathrm{th}}$ distributions for the heavy-ion exposures performed in this study. The solid lines indicate experimental data, while the markers indicate simulated results. The error bars on the simulated points represent statistical uncertainty and read-noise. Inset to figure~\ref{fig:Results} (Left) are the model parameters $b_e$ and $b_i$ used in the simulation. Interestingly, the best agreement between simulation and experiment was obtained with $b_i=0$ for all ion species considered. Within the assumptions of the model, this suggests that the observed trapped-charge discharge is dominated by secondary electrons rather than by the primary heavy ion directly. This result is also consistent with an electron-mediated discharge mechanism. However, because holes generated in the oxide are not explicitly included in the present model, this interpretation should be viewed as suggestive rather than conclusive. The measured $\Delta V_{\mathrm{th}}$ distributions are consistent with previously published HIMoC results, in which ions with larger atomic number, $Z$, and larger energy per nucleon produced distributions with longer negative-shift tails~\cite{HIMoC_Paper}. Very little difference is observed between the two $^{22}$Ne exposures, despite the exposure times differing by a factor of $\sim33$. This is the expected response for a dosimeter, for which the total induced signal should depend primarily on the accumulated dose rather than the time over which the dose is delivered. A large peak near $\Delta V_{\mathrm{th}}=0$ is observed in all distributions. This peak is associated with background bits that did not undergo radiation-induced discharge, but which exhibit small apparent threshold-voltage shifts due to readout noise. As discussed above, repeated readout of an unirradiated device produced an approximately Gaussian read-noise distribution with measured values of $\mu_{\mathrm{RN}}=-12.1~\mathrm{mV}$ and $\sigma_{\mathrm{RN}}=39.0~\mathrm{mV}$. For the simulations performed, the read-noise parameters were allowed to vary to match each device, reflecting device-to-device differences in the readout noise distribution. Table~\ref{tab:parameters} summarizes the read-noise values applied for each simulation, as well as the mean secondary-electron multiplicity and mean LET extracted from Geant4 for each heavy ion.

It is observed that the simulation workflow produces very accurate results for the $^{14}$N irradiation, while the $^{22}$Ne simulation produces less comparative agreement in the far tail regions of the distribution, and the $^{40}$Ar simulation under-populates low $\Delta V_{\mathrm{th}}$ bins and over populates high $\Delta V_{\mathrm{th}}$ bins. It is proposed that mismodeling of bins between $-500\;\mathrm{mV}<\Delta V_{\mathrm{th}}<0\;\mathrm{mV}$ are a result of very low energy electrons which are poorly modeled. In particular, any electrons with energy beneath the production threshold set in the \textit{Geant4-MicroElec} package will not be accounted for, which is precisely the range which would contribute most to filling those bins. Ultimately, these bins are not included in the signal measurement and are therefore less impactful overall, as discussed below. It is further proposed that the overpopulated bins in the tails of the $^{22}$Ne and $^{40}$Ar distributions might originate from the modeling of energy deposition by Geant4 in very thin sensitive volumes (approximately 15nm in thickness). Such thin volumes would introduce a larger variance in energy deposition, which might manifest as an overestimation of total energy deposition and therefore an overpopulation of those bins. 

In order to select radiation-induced signal bits while reducing the contribution from readout noise, a cutoff value, $C_V$, was imposed. Bits with threshold-voltage shifts more negative than $C_V$ were classified as signal bits. In this study, we define a custom metric referred to as the \textit{summed threshold-voltage shift}, $\Sigma \Delta V_{\mathrm{th}}$, given by the sum of $\Delta V_{\mathrm{th}}$ over all bits satisfying $\Delta V_{\mathrm{th}} < C_V$. Figure~\ref{fig:Results} (Right) shows $\Sigma \Delta V_{\mathrm{th}}$ for each exposed HIMoC device, using the cutoff value $C_V=-500~\mathrm{mV}$. The horizontal axis is the product of the ion fluence, the mean LET extracted from Geant4, and the active area of the device. Because all HIMoC devices tested have identical dimensions, this quantity is proportional to the total energy deposited in the sensitive layer and therefore to the absorbed dose. Across the ion species and fluences considered, $\Sigma \Delta V_{\mathrm{th}}$ exhibits a linear response with this dose-like quantity, demonstrating the potential of HIMoC as a passive heavy-ion dosimeter.

\begin{table}[tbp]
\centering
\caption{Summary of heavy-ion and secondary-electron simulation quantities.}
\label{tab:parameters}
\begin{tabular}{|c|c|c|c|c|}
\hline
Ion Species &
\begin{tabular}[c]{@{}c@{}}$\mu_{\mathrm{RN}}$\\ $\mathrm{[mV]}$\end{tabular} &
\begin{tabular}[c]{@{}c@{}}$\sigma_{\mathrm{RN}}$\\ $\mathrm{[mV]}$\end{tabular} &
\begin{tabular}[c]{@{}c@{}}Mean electron\\ multiplicity\end{tabular} &
\begin{tabular}[c]{@{}c@{}}Mean LET\\ $\mathrm{[MeV/cm]}$\end{tabular} \\
\hline
24.8 MeV/u $^{14}$N  & -24.2 & 54.6 & 2,920  & 0.66 \\
\hline
24.8 MeV/u $^{22}$Ne & -30.2 & 51.4 & 5,743  & 1.36 \\
\hline
24.8 MeV/u $^{40}$Ar & -24.2 & 54.6 & 18,627 & 4.74 \\
\hline
\end{tabular}
\end{table}

\subsection{Implications for Charge-Trapping NVM in Radiation Environments}

An important outcome of this work is the development of a TCAD-based modeling approach for simulating arbitrary trapped-charge configurations in the charge-trapping layers of non-volatile memory (NVM) devices. Although developed here for HIMoC, this capability is also relevant to understanding the radiation response of NVM technologies in high-radiation environments, particularly space. It has recently been proposed that space-based data centers could help alleviate resource and energy constraints associated with terrestrial data centers~\cite{datacenters}. Charge-trapping NVM devices that use $\mathrm{Si}_{3}\mathrm{N}_{4}$, including SONOS configurations in which charge is distributed evenly across the full channel rather than localized into two bits like in NROM, have also been proposed as promising candidates for improved radiation resistance in NVM applications~\cite{SONOS}. Therefore, if future space-based computing systems incorporate large numbers of SONOS or related charge-trapping NVM devices, accurate modeling of their radiation response will be impactful for understanding and mitigating potential data-corruption mechanisms. 

NVM devices have also been proposed for neuromorphic computing architectures, including machine-learning (ML) applications~\cite{Neuromorphic_computing}. In such architectures, the weights and biases of trained ML models can be encoded by the amount of charge stored in the charge-trapping regions of NVM devices. Heavy-ion irradiation can therefore lead to degraded inference performance through radiation-induced charge loss~\cite{ML_effects}. The ability to predict how specific heavy-ion species and energies alter trapped-charge configurations could therefore enable quantitative prediction of model degradation in new radiation environments.

\subsection{Future Work}

Several future research directions are motivated by this work. Further experimental characterization of HIMoC is needed to verify linear response for ions of the same species across varying energies, and to evaluate device response at dose levels relevant to astronaut exposure and other space-radiation environments. In addition, DEVSIM is capable of performing time dependent simulations which could directly contribute to better deducing the exact dynamics of NVM under heavy ion irradiation and therefore determine the most impactful charge depletion mechanism. Such studies may help improve both the radiation hardness of charge-trapping NVM technologies and their use as passive radiation sensors.

\section{Conclusion}\label{sec:Conclusion}

We have further characterized the heavy-ion-induced response of HIMoC devices using exposures to 24.8 MeV/u $^{14}$N, $^{22}$Ne, and $^{40}$Ar ions. Across the ion species and fluences tested, the $\Sigma \Delta V_{\mathrm{th}}$ response was found to scale linearly with a dose-like quantity proportional to fluence, LET, and active detector area. This result supports the use of HIMoC as a passive heavy-ion dosimeter capable of recording radiation interactions without active power during exposure. A full simulation workflow was developed to model the HIMoC response. Geant4 simulations were used to model primary-ion energy deposition, secondary-electron production, and secondary-electron energy deposition. Three-dimensional TCAD simulations in DEVSIM were then used to extract the relationship between localized charge loss in the NROM charge-trapping layer and the resulting $\Delta V_{\mathrm{th}}$. These components were combined into a sector-based simulation workflow capable of producing simulated $\Delta V_{\mathrm{th}}$ distributions for direct comparison with experiment. The best agreement between simulation and measurement was obtained when the heavy-ion contribution to the Gaussian charge-loss radius was set to zero, indicating that the measured response is likely dominated by secondary-electron-mediated discharge rather than by direct discharge from the primary ion. This interpretation is consistent with the importance of secondary-electron modeling observed in the simulation workflow, although additional studies are required to determine the exact charge-depletion mechanism. Future work will focus on further experimental characterization across a broader range of ion species, energies, fluences, and dose rates, including exposures relevant to astronaut and space-electronics environments. Additional time-dependent TCAD simulations may also help distinguish between possible charge-loss mechanisms and improve predictive models for both HIMoC dosimetry and the radiation response of charge-trapping NVM devices more broadly.

\section{Acknowledgments}

This work was supported in part by the Texas A\&M University Cyclotron Institute  (DOE DE-FG02-93ER40773, NSF PHY-2051072, NSF PHY-2447482). We would like to thank Juan Sanchez for his helpful discussions related to DEVSIM.

%\newpage

\end{document}